# Physical origin of Davydov splitting and resonant Raman spectroscopy of Davydov components in multilayer MoTe$_2$

Q. J. Song, [1, 3, †] Q. H. Tan, [2, †] X. Zhang, [2] J. B. Wu, [2] B. W. Sheng, [1] Y. Wan, [1, 3] X. Q. Wang, [1, 3] L. Dai, [1, 3, ‡] and P. H. Tan [2, *]

[1] *State Key Lab for Mesoscopic Physics and School of Physics, Peking University, Beijing 100871, China.*

[2] *State Key Laboratory of Superlattices and Microstructures, Institute of Semiconductors, Chinese Academy of Sciences, Beijing 100083, China.*

[3] *Collaborative Innovation Center of Quantum Matter, Beijing 100871, China.*

† These authors contributed equally to this work.

‡ lundai@pku.edu.cn

* phtan@semi.ac.cn

## ABSTRACT

We systematically study the high-resolution and polarized Raman spectra of multilayer (ML) MoTe$_2$ . The layer-breathing (LB) and shear (C) modes are observed in the ultralow-frequency region, which are used to quantitatively evaluate the interlayer coupling in ML MoTe$_2$ based on the linear chain model, in which only the nearest interlayer coupling is considered. The Raman spectra on three different substrates verify the negligible substrate effect on the phonon frequencies of ML MoTe$_2$. Ten excitation energies are used to measure the high-frequency modes of N-layer MoTe$_2$ (NL MoTe$_2$; N is an integer). Under the resonant excitation condition, we observe N–dependent Davydov components in ML MoTe$_2$, originating from the Raman-active $A_1'(A_{1g}^2)$ modes at ~172 cm$^{-1}$. More than two Davydov components are observed in NL MoTe$_2$ for N > 4 by Raman spectroscopy. The N-dependent Davydov components are further investigated based on the symmetry analysis. A van der Waals model only considering the nearest interlayer coupling has been proposed to well understand the Davydov splitting of high-frequency $A_1'(A_{1g}^2)$ modes. The different resonant profiles for the two Davydov components in 3L MoTe$_2$ indicate that proper excitation energy of ~1.8 − 2.2 eV must be chosen to observe the Davydov splitting in ML MoTe$_2$. Our work presents a simple way to identify layer number of ultrathin MoTe$_2$ flakes by the corresponding number and peak position of Davydov components. Our work also provides a direct evidence from Raman spectroscopy of how the nearest van der Waals interactions significantly affect the frequency of the high-frequency intralayer phonon modes in multilayer MoTe$_2$ and expands the understanding on the lattice vibrations and interlayer coupling of transition metal dichalcogenides and other two-dimensional materials.

## I. INTRODUCTION

Transition metal dichalcogenides (TMDs) with the formula MX$_2$, where M is a transition metal (Mo, W, and so on) and X is a chalcogen (S, Se, or Te), have attracted much attention due to their many interesting physical properties, such as direct band gap for monolayer [1,2], valley polarization [3–5], valley Hall effect [6], tightly bonded trions [7], second-harmonic generation [8–11], etc. Compared with the commonly reported TMDs, such as MoS$_2$, MoSe$_2$, WS$_2$, and WSe$_2$ [1,2,12,13], MoTe$_2$ has a narrower direct band gap (~1.1 eV) in one monolayer (1L) [14], and is an ideal candidate material for infrared optoelectronic devices. Additionally, stronger spin-orbit coupling and thermally induced structural phase translation have been revealed in few-layer MoTe$_2$ [14–17].

As a fast and nondestructive characterization technique, Raman spectroscopy has been extensively used to provide structural and electronic information of layered TMDs [13]. Specifically, ultralow-frequency Raman spectroscopy has been used to investigate the interlayer and interface coupling of layered materials, such as multilayer graphene (MLG), and multilayer (ML) MoS$_2$ and WSe$_2$ [13,18–21]. The linear chain model has been used to understand the interlayer vibrations in those layered materials [13,18,20,21] where the interaction between layered materials and substrate is neglected. Also, the second-nearest layer-breathing interlayer coupling has been found to exist in MLGs [21]. Moreover, a recent study revealed that a substrate-induced interface mode has been observed in Bi$_2$Te$_3$/Bi$_2$Se$_3$ two-dimensional crystals on mica and SiO$_2$ substrates [22]. It is necessary to experimentally confirm whether the second-nearest layer-breathing interlayer coupling and such substrate effect exist in the TMD system, such as MoTe$_2$. Up to now, only a few works about few-layer MoTe$_2$ have been reported. Yamamoto et al. uncovered the strong enhancement of the bulk inactive $B_{2g}^1$ mode in few-layer MoTe$_2$ [17]. Guo et al. studied the resonant mechanism of second-order Raman modes [23]. However, the resonant behavior of the first-order Raman modes, such as the $A_1'(A_{1g}^2)$ mode in multilayer MoTe$_2$, is not clear. Most recently, Froehlicher et al.reported the observation of Davydov splitting in multilayer MoTe$_2$ [24], which is quantitatively described by a force constant model, including interactions up to the second nearest neighbor and surface effects as fitting parameters. However, it is well known that the frequency of the shear (C) and layer-breathing (LB) modes in TMDs can be well described by the linear chain model that only the nearest interlayer coupling is taken into account [13,18,19]. Van der Waals (vdW) interlayer interactions are much weaker than the strength of the covalent bonds between the atoms within each layer [13,18], and thus with respect to the C and LB modes, it is unclear that why the second nearest neighbor and surface effects [24] should be included to understand the small Davydov splitting of the high-frequency optical modes in multilayer $MoTe_2$. Therefore, more works are needed to reveal the intrinsic origin of Davydov splitting in few-layer $MoTe_2$ and how the vdW interactions affect the frequency of the high-frequency intralayer phonon modes in multilayer TMDs.

In this work, we studied the high-resolution Raman spectra of N-layer $MoTe_2$ (NL $MoTe_2$; N is an integer) in both ultralow-frequency and high-frequency regions. In the ultralow-frequency region, we have observed the branch of LB modes with the lowest frequency and the branch of C modes with the highest frequency. The LB and C modes are identified by the parallelly and perpendicularly polarized spectra, and their frequencies are in accordance with those calculated by the linear chain model (LCM), in which the interaction between NL $MoTe_2$ and substrate has been ignored. Then the force constants are calculated to investigate the interlayer coupling. By comparing the Raman spectra on three different substrates, we demonstrate that the substrate has little influence on the vibration modes. In the high-frequency region, we observed the Davydov components of the $A_1^{'}(A_{1g}^2)$ mode in ML $MoTe_2$ under the resonant excitation condition. The number and peak position of Davydov components are dependent on the layer number of NL $MoTe_2$, which also provides a simple way to identify the layer number of ultrathin $MoTe_2$ flakes. The N-dependent Davydov components are further investigated based on the symmetry analysis. It is noteworthy that the Davydov splitting can be well understood by a simple van der Waals model, in which only the nearest interlayer coupling is considered. The different resonant profile of the Davydov components in 3L MoTe$_2$ clarifies the importance of the laser excitation energy in the observation of the Davydov splitting in NL MoTe$_2$ (N > 2). Our results are helpful in the investigation on the interlayer coupling in other layered materials by Davydov splitting.

## II. EXPERIMENT

The NL MoTe$_2$ samples were prepared by mechanical exfoliation from the bulk material (2D Semiconductors, Inc.) and then deposited on various substrates, including 300-nm SiO$_2$/Si, quartz, and sapphire. All the results presented in this paper are obtained from the samples on 300-nm SiO$_2$/Si substrates unless otherwise stated.

Raman and photoluminescence (PL) measurements were performed in a commercial micro-Raman setup (Horiba Jobin Yvon HR800) under the backscattering geometry. The Raman system is equipped with a liquid nitrogen–cooled chargecoupled detector (CCD), a 100× objective lens (numerical aperture of 0.90), and several gratings. The laser excitation energies (Elas) are 1.58 and 1.71 eV from a Ti:sapphire laser; 1.96, 2.08, and 2.28 eV from a He-Ne laser; 1.83, 2.18, 2.34, and 2.41 eV from a Kr+ laser; and 2.54 eV from an Ar+ laser. The ultralow-frequency Raman spectra were obtained down to ± 5 cm$^{-1}$ by combining three volume Bragg grating filters (BragGrate notch filters, OptiGrate Corp.) into the Raman system to effectively suppress the Rayleigh signal [25]. Each BragGrate notch filter has optical density 3 and full width at half maximum (FWHM) of 5–10 cm$^{-1}$ [20,21]. We kept the laser power below 0.4 mW to avoid the heating effect to the samples.

## III. RESULTS AND DISCUSSION

## A. Ultralow frequency shear and layer-breathing modes in multilayer MoTe$_2$

Two typical optical images of the samples are shown in Fig. 1(a). The layer number of MoTe$_2$ flakes was identified by PL emission and atomic force microscopy (AFM). Compared with indirect band gap NL MoTe$_2$ (N≥2), 1L MoTe$_2$, which is a direct band-gap semiconductor induced by the absence of interlayer coupling [1,2,14], exhibits the strongest PL intensity and the largest band-gap energy. The 1L MoTe$_2$ can be further identified by the lack of ultralow-frequency modes [26], as discussed later. The layer number of NL MoTe$_2$ (N≥2) can then be obtained from its thickness difference from that of 1L MoTe$_2$, where the thickness of one layer in NL MoTe$_2$ is about 0.7 nm, as shown in Fig. 1(b). Notably, for the sample of 1L and other-layer MoTe$_2$ in Fig. 1(b), the minimum height between sample and substrate is about 4.7 nm. This is reasonable in consideration of the instrumental offset between the samples and substrate [13,27,28]. The typical PL spectra of 1L and 3L MoTe$_2$ on 300-nm SiO$_2$/Si substrate are shown in Fig. 1(c), where the two peaks around 0.93 eV and 0.98 eV come from the Si substrate.

The ultralow-frequency Raman spectra of 1L–6L MoTe$_2$ under Elas of 2.28 eV are shown in Fig. 2(a). The two spikes at around ± 4.5 cm$^{-1}$ come from the Brillouin scattering of the longitudinal acoustic (LA) mode of Si substrate [18]. Their intensities decrease with increasing N and their frequencies almost keep constant. No ultralow-frequency peak appears in 1L MoTe$_2$ owing to the absence of interlayer interaction [13,18,25]. For 2L MoTe$_2$, there are two peaks located at ∼18.8 and ∼27.8 cm$^{-1}$. For 3L MoTe$_2$, two peaks are observed at ∼19.6 and ∼23 cm$^{-1}$. For 4L–6L MoTe$_2$, three modes are observed. According to the symmetry analysis [13,18], there are N–1 LB modes and N–1 twofold degenerate C modes in NL MoTe$_2$ (N > 1), which can be denoted as $C_{N,N-j}$ and $LB_{N,N-j}$, respectively, where j = N–1, N–2,..., 1, and $C_{N,1}$ and $LB_{N,1}$ are the C and LB modes with the highest frequencies, respectively. In order to distinguish the LB and C modes, we measured the Raman spectra of 2L–6L MoTe$_2$ under parallel ($\bar{z}(xx)z$) and perpendicular ($\bar{z}(xy)z$) polarization configurations, as shown in Fig. 2(b). Based on their Raman tensors [13], the LB modes only appear in $\bar{z}(xx)z$ polarization configuration, while the C modes can be observed under both $\bar{z}(xx)z$ and $\bar{z}(xy)z$ polarization configurations. Therefore, we assign the modes observed in the $\bar{z}(xy)z$ polarization configuration as the C modes, and the rest of ultralow-frequency modes as the LB modes.

Because the C and LB modes originate from the relative motions of the rigid monolayer planes themselves in ML MoTe$_2$, each rigid monolayer plane (a Mo plane sandwiched by two Te planes) can be treated as a ball to analyze the atomic displacements of the interlayer modes, i.e., the so-called linear chain model (LCM) [13,18,25,29]. When only the interlayer coupling between the nearest-neighbor layers is considered, the eigenfrequencies in NL MoTe$_2$ for the C and LB modes can be expressed as follows [21]:

$$\omega(C_{N,N-j}) = \sqrt{2}\omega(C_{21})\sin(\frac{j\pi}{2N}), \qquad (1)$$

$$\omega(LB_{N,N-j}) = \sqrt{2}\omega(LB_{21})\sin(\frac{j\pi}{2N}), \qquad (2)$$

where ω(C21) and ω(LB21) are the frequencies of the C and LB modes in 2L MoTe$_2$, respectively. The branches of j = N–1 and j = 1 are observed for the C and LB modes, respectively. If we denote $m_{Mo}$ ($m_{Te}$) as the mass per unit area for Mo (Te) atom planes and $\alpha_{TeTe}^{\parallel}$ ($\alpha_{TeTe}^{\perp}$) as the parallel (perpendicular) component of the force constant per unit area between two nearest Te atoms planes in two adjacent layers, then $\omega(C_{21}) = \frac{1}{\sqrt{2}\pi c}\sqrt{\frac{\alpha_{TeTe}^{\parallel}}{m_{Mo}+2m_{Te}}}$ and $\omega(LB_{21}) = \frac{1}{\sqrt{2}\pi c}\sqrt{\frac{\alpha_{TeTe}^{\perp}}{m_{Mo}+2m_{Te}}}$. The calculated frequencies of the C and LB modes based on the LCM are summarized in Figs. 2(c) and 2(d), respectively. The corresponding experimental data are also summarized in Figs. 2(c) and 2(d). The theoretical and experimental data are in good agreement with each other, demonstrating that the second nearest C and LB interlayer coupling can be ignored. Using the detected $\omega(C_{21})$ and $\omega(LB_{21})$, we can obtain $\alpha_{TeTe}^{\perp}$ and $\alpha_{TeTe}^{\parallel}$ to be about $9.12\times10^{19}$ N/m³ and $4.25\times10^{19}$ N/m³, respectively. The force constant $\alpha_{ss}^{\perp}$ and $\alpha_{ss}^{\parallel}$ in MoS$_2$ is about $8.90 \times 10^{19}$ N/m³ and $2.82 \times 10^{19}$ N/m³ [18]. It shows that the force constant in MoTe$_2$ is slightly larger than in MoS$_2$, indicating small difference of interlayer coupling strength between MoTe$_2$ and MoS$_2$.

**B. Negligible substrate effect on the Raman modes in multilayer MoTe$_2$**

It is worth noting that the intensity of the LB mode with the lowest frequency in ML-MoTe$_2$ is much stronger than that of the C mode, as shown in Fig. 2(a). This phenomenon is opposite to the reported results for ML-MoS$_2$ and WSe$_2$ [18,19]. The interaction between $N$L-MoTe$_2$ and the substrate may influence the intensity of out-of-plane LB modes [22]. In order to explore the substrate effect, we measured the Raman spectra of ML MoTe$_2$ on both quartz and sapphire substrates. Figure 3 shows the representative Raman spectra of 6L MoTe$_2$ on the three different substrates. Each spectrum has been normalized in intensity to the lowest-frequency LB mode in 6L MoTe$_2$. The peaks located at ~521 and ~417 cm$^{-1}$ come from 300-nm SiO$_2$/Si and sapphire substrates, respectively. The $A_{1g}^2$, $E_g$ and $A_{1g}^1$ modes observed at ~172, ~233, and ~289 cm$^{-1}$, respectively. Two LB modes and one C mode are also observed in the ultralow-frequency region. The relative intensity between any two modes in 6L MoTe$_2$ stays almost constant on the three different substrates. Furthermore, the frequencies of the Raman modes in both ultralow-frequency and high-frequency regions remain almost unchanged. This reveals that the coupling between the substrate and ML MoTe$_2$ can be ignored in the experiments, confirming the assumption for the LCM in the calculation for $\omega_{(CN,N-j)}$ and $\omega_{(LBN,N-j)}$ in Eqs. (1) and (2). This rules out the substrate effect on the strong intensity of LB modes. The strong intensity of the j = 1 branch for the LB mode may result from its strong electron-phonon coupling in ML MoTe$_2$ [25]. Because the interaction between substrate and ML MoTe$_2$ can be ignored, the frequency of C and LB modes is a simple and reliable way to identify the layer number of ML-MoTe$_2$.

## C. Davydov splitting in Multilayer MoTe$_2$

Now, we focus on the high-frequency Raman modes in NL MoTe$_2$. There are six optical modes in 1L MoTe$_2$, in which three high-frequency optical modes ($A_1'$, two-fold degenerate $E'$ and $E''$ modes) are Raman active and additional high-frequency $A_2''$ mode are infrared (IR) active [13]. Figure 4(a) shows the Raman spectra of NL MoTe$_2$ (N = 1–6) and bulk MoTe$_2$ in the high-frequency region excited by Elas of 2.28 eV. Only two vibration modes ($A_1'$ and $E'$) exist in 1L-MoTe$_2$, which correspond to the $A_{1g}$ (~173 cm$^{-1}$) and $E_{2g}^1$ (~233 cm$^{-1}$) modes in bulk MoTe$_2$, respectively. The $E''$ mode is not observed under the backscattering configuration based on its Raman tensor [13]. For NL MoTe$_2$ (N >2), corresponding to the $A_1'$ and $E'$ modes in 1L MoTe$_2$, the $A_1'$ and $E'$ modes in odd number layers (ONL-) MoTe$_2$ and the $A_{1g}^2$ and $E_g$ modes in even number layers (ENL-) MoTe$_2$ are observed, respectively. An additional mode at ~291 cm$^{-1}$ is observed in NL MoTe$_2$, which is assigned as the $A_2''$ mode in ONL-MoTe$_2$ and the $A_{1g}^1$ mode in ENL MoTe$_2$, respectively. This mode corresponds to the Raman-inactive $B_{2g}^1$ mode in bulk MoTe$_2$ [17]. Here, the different denotations of modes in bulk, ONL, and ENL MoTe$_2$ originate from their symmetry difference [13].

It seems that the $A_1'(A_{1g}^2)$ mode of 3L–5L MoTe$_2$ exhibits multiple peaks excited by Elas of 2.28 eV in Fig. 4(a). To clearly reveal this spectral feature, Elas of 1.96 eV is used to excite the Raman spectra of 1L–6L MoTe$_2$, and the corresponding Raman spectra are depicted in Fig. 4(b). The relative intensity of the $A_1'$ mode to the $E'$ mode in 1L MoTe$_2$ is significantly enhanced because Elas of 1.96 eV is close to the energy of the B exciton (~2.0 eV). The $A_1'$ mode in 1L MoTe$_2$ (~171.6 cm$^{-1}$) blueshifts to ~171.9 cm$^{-1}$ of the $A_{1g}^2$ mode in 2L MoTe$_2$. Interestingly, the corresponding modes in NL MoTe$_2$ clearly exhibit multiple components for N > 2. One can clearly see two peaks in 3L MoTe$_2$ (~168.9 and 172.2 cm$^{-1}$) and 4L MoTe$_2$ (~169.8 and 172.5 cm$^{-1}$), and three peaks in 5L MoTe$_2$ (~168.6, 170.7, and 172.8 cm$^{-1}$) and 6L MoTe$_2$ (~168.9, 171.6, and 172.9 cm$^{-1}$). The average frequency of the multiple peaks of 3L–6L MoTe$_2$ is close to the $A_1'$ frequency of 1L MoTe$_2$, suggesting that the multiple peak are the out-of-plane $A_1'(A_{1g}^2)$ modes in NL MoTe$_2$, which is derived from the $A_1'$ mode in 1L MoTe$_2$. We further measured the polarized Raman spectra of 1L–4L MoTe$_2$ in the high-frequency region under 1.96 eV excitation, as shown in Fig. 4(c). All the multiple peaks appear in the $\bar{z}(xx)z$ polarization configuration, but completely vanish in the $\bar{z}(xy)z$ polarization configuration, which is consistent with the Raman selection rule and the Raman tensor of the $A_1'(A_{1g}^2)$ modes [13]. The polarization Raman result confirms that these multiple components are closely related to the out-of-plane $A_1'(A_{1g}^2)$ modes in NL MoTe$_2$. In order to reveal the evolution of multiple peaks inNL MoTe$_2$, we zoom in on the spectral region of the $A_1'(A_{1g}^2)$ modes in Fig. 4(b), as shown in Fig. 5(a). The spectra are normalized to their strongest peak and are offset for clarity. There exist three sets (R$_1$, R$_2$, and R$_3$) of Raman peaks associated with the $A_1'(A_{1g}^2)$ mode in 1L–6L MoTe$_2$, as indicated by the three dashed lines. Figure 5(b) shows the layer number dependence of the frequencies for the three sets of the $A_1'(A_{1g}^2)$ modes. With increasing the layer number, clear stiffening can be observed for each set of Raman peaks.

In comparison to 1L MoTe$_2$, there exists additional interlayer coupling in ML MoTe$_2$, and thus the A 1 mode in 1L MoTe$_2$ will split into N modes in NL MoTe$_2$ (N > 1). These modes can be expressed as $\frac{N+1}{2}A_1' + \frac{N-1}{2}A_2''$ for ONL-MoTe$_2$ and $\frac{N}{2}A_{2u} + \frac{N}{2}A_{1g}^2$ for ENL MoTe$_2$ [13,30], where the $A_1'$ and $A_{1g}^2$ modes are Raman active and the $A_2''$ and $A_{2u}$ modes are infrared active. The atomic displacements of the corresponding modes in 1L–6L MoTe$_2$ are depicted in Fig. 5(c) along with the symmetry denotations, where R$_j$ and IR$_j$ (j = 1, 2, or 3) are used to distinguish the Raman- and infrared-active modes in NL MoTe$_2$ with the same symmetry, respectively. It is evident that the number of the observed Raman modes at ~170 cm$^{-1}$ is exactly equal to that of Raman-active $A_1'(A_{1g}^2)$ modes in NL MoTe$_2$, suggesting that the observed multiple Raman peaks in 3L–6L MoTe$_2$ are the corresponding Raman-active $A_1'(A_{1g}^2)$ modes.

The frequency of the $A_1'(A_{1g}^2)$ modes in ML MoTe$_2$ is dependent on the interlayer coupling in adjacent layers, i.e., the coupling between two nearest Te atoms in adjacent layers if only the nearest coupling is considered. Once the nearest Te atoms in adjacent layers vibrate out of phase, the additional vdW interaction between the nearest Te atoms in adjacent layers will raise the frequency of the Raman mode with respect to the mode whose nearest Te atoms in adjacent layers vibrate in phase. Consequently, the frequency of the $A_1'(A_{1g}^2)$ modes in ML MoTe$_2$ is sensitive to the number of out-of-phase vibrations of the nearest Te atoms in adjacent layers. For example, in 3L MoTe$_2$, there are two, one, and zero out-of-phase vibrations between the nearest Te atoms in adjacent layers for the $A_1'$, $A_2''$ (IR$_1$), and $A_{1g}^2$ modes, respectively, as illustrated by the corresponding atomic displacements in Fig. 5(c); thus the $A_1'(R_1)$ mode is with the highest frequency and the $A_1'(R_2)$ mode is with the lowest frequency in 3L MoTe$_2$. This can also be applied to other NL MoTe$_2$. The $A_1'$ or $A_1^2$ modes are with the highest frequency for each ML MoTe$_2$ in Fig. 5(c) because all the nearest Te atoms in adjacent layers vibrate out of phase.

When two 1L MoTe$_2$ are combined together to be a 2L MoTe$_2$, each optical mode in 1L MoTe$_2$ will split into the corresponding two modes in 2L MoTe$_2$, of which the two nearest Te atoms in adjacent layers in the unit cell vibrate in phase in one mode and out of phase in the other mode. For example, the A$_{2u}$(IR$_1$) and $A_{1g}^2$(R$_1$) modes in 2L-MoTe$_2$ are derived from the $A_1'$(R$_1$) mode in 1L-MoTe$_2$, as indicated in Fig. 5(c). This is also true in bulk MoTe$_2$ because its unit cell is the same as that of 2L MoTe$_2$. The frequency difference between the two modes is determined by the vdW interaction between two layers in the unit cell, which is well known as Davydov splitting in bulk and 2L TMD [13,31,32]. Based on the symmetry analysis for bulk and 2L MoTe$_2$, only one of the Davydov doublets can possibly be Raman active, and thus it is difficult to observe the Davydov doublets by Raman spectroscopy in bulk and 2L MoTe$_2$. In fact, the general Davydov splitting is known as the splitting of bands in the electronic or vibrational spectra of crystals due to the presence of more than one (interacting) equivalent molecular entity in the unit cell [33]. Indeed, three and four equivalent entities can be found in many systems and the corresponding Davydov components have been observed [34–36]. Thus, the concept of Davydov splitting related with bulk and 2L TMDs can be extended for NL MoTe$_2$ (*N*>2), for which there exist N equivalent entities in its unit cell for Davydov splitting. Each equivalent (isolated) entity is a Mo atom sandwiched by two Te atoms in the unit cell of 1L MoTe$_2$. Each mode in 1L-MoTe$_2$ (e.g., the $A_1'$ mode) can derive into *N* corresponding modes in *N*L-MoTe$_2$ (e.g., the $A_1'$ and $A_2''$ modes in ONL-MoTe$_2$, or the $A_{1g}^2$ and A$_{2u}$ modes in ENL-MoTe$_2$) resulting from the different coupling cases of the NMoTe$_2$ layers. Only one mode ccorresponds to the uncoupled entities in which all the nearest Te atoms in adjacent layers vibrate in phase, e.g., the $A_1'$ (R$_2$) mode in 3L-MoTe$_2$ and the A$_{2u}$ (IR$_2$) mode in 4L-MoTe$_2$. The other N–1 modes correspond to the N–1 coupled entities in which at least one pair of nearest Te atoms in adjacent layers vibrates out of phase. The out-of-phase vibrations between nearest Te atoms in adjacent layers of the coupled entities will result in a frequency different from the uncoupled entities, and Davydov components are formed in NL MoTe$_2$. Therefore, each optical mode in 1L MoTe$_2$ can correspond to N corresponding Davydov components in NL MoTe$_2$ (N > 1), which can be directly observed by Raman spectroscopy once at least two Davydov components are Raman active. Indeed, one couple of Davydov components in NL MoSe$_2$ has been observed for N = 3, 4, and 5 by Tonndorf et al. [37]. Figure 5(a) shows that the number of the observed modes in NL MoTe$_2$ at ~171 cm$^{-1}$ is exactly equal to that of Raman-active $A_1^{'}$($A_{1g}^2$) modes, suggesting that these observed modes are Davydov components in $N$L-MoTe$_2$ corresponding the $A_1^{'}$ mode in 1L-MoTe$_2$.

**D. The vdW model for Davydov splitting in Multilayer MoTe$_2$**

We can estimate Davydov splitting between two Davydov components in NL MoTe$_2$. It is well known that two identical coupled entities have vibrational frequencies given by ω$_0$ and ω$_c$, where ω$_0$ is the frequency of the isolated entity and that of two uncoupled entities when the two entities vibrate in phase; ω$_c$ is the frequency of two coupled entities in which they vibrate out of phase. If Δω is the coupling frequency between two coupled entities, the three frequencies have the relation of $\omega_c^2 = \omega_0^2 + \Delta\omega^2$ [38]. We can extend this concept to the $A_{1g}^2$ and A$_{2u}$ modes in 4L MoTe$_2$ as an example if the interlayer forces are exclusively of the vdW interactions, which is denoted as the vdW model. The interlayer coupling between the $A_1^{'}$ mode of 1L MoTe$_2$ generates two $A_{1g}^2$ and two A$_{2u}$ modes and forms Davydov components in 4L MoTe$_2$. The A$_{2u}$(IR$_2$) mode is the uncoupled entities with a frequency of ω$_0$, while the $A_{1g}^2$ (R$_2$), A$_{2u}$(IR$_1$), and $A_{1g}^2$ (R$_1$) modes are three coupled entities with frequencies of ω$_{c3}$, ω$_{c2}$, and ω$_{c1}$, respectively. There are one, two, and three pairs of out-of-phase vibrations between nearest Te atoms in adjacent layers for the $A_{1g}^2$, $A_{2u}(IR_1)$, and $A_{1g}^2$ ($R_1$) modes, respectively, as demonstrated in Fig. 6(a). Thus, $\omega_{c1}<\omega_{c2}<\omega_{c3}$. Based on the vdW model, the vdW interlayer coupling results in the frequency difference between the four Davydov components in 4L MoTe$_2$. Because the atomic displacements of the $A_{1g}^2$ and $A_{2u}$ modes are perpendicular to the basal plane, the interlayer LB coupling is responsible for the frequency difference between two Davydov components. Indeed, as depicted in Fig. 6(a), the LB$_{43}$, LB$_{42}$, and LB$_{41}$ modes in 4L MoTe$_2$ exhibit the same number of out-of-phase vibrations between nearest Te atoms with respect to the three coupled $A_{1g}^2$ ($R_2$), $A_{2u}(IR_1)$, and $A_{1g}^2$ ($R_1$) modes, respectively. Therefore, the interlayer coupling strength within the coupled $A_{1g}^2$ ($R_2$), $A_{2u}(IR_1)$, and $A_{1g}^2$ ($R_1$) modes is directly reflected by the frequency of the LB$_{43}$, LB$_{42}$, and LB$_{41}$ modes, respectively, and the frequency of the LB$_{4j}$ modes (j = 1, 2, 3) is the coupling frequency ($\Delta\omega_j$) for the corresponding coupled modes. The frequency ($\omega_{cj}$) of each coupled Davydov component and the corresponding coupling frequency ($\Delta\omega_j$) will follow the relation of $\omega_{cj}^2 - \Delta\omega_j^2 = \omega_0^2$ (j = 1,2,3) for 4L MoTe$_2$, which can be used to estimate Davydov splitting or the mode frequency of Davydov component in 4L MoTe$_2$. Figure 6(a) demonstrates the schematic diagram of the vdW model for Davydov splitting in 4L MoTe$_2$, where the LA mode frequency can be regarded as the coupling frequency (zero) for the uncoupled $A_{2u}(IR_1)$ modes. For instance, the Davydov splitting between the $A_{1g}^2$ ($R_2$) and $A_{1g}^2$ ($R_1$) modes is calculated to be 3.1 cm$^{-1}$, which is in good agreement with the experimental result (2.7 cm$^{-1}$). The above description of the vdW model for the Davydov splitting in 4L MoTe$_2$ can be extended to Davydov splitting of any layered materials associated with the shear or layer-breathing couplings if their interlayer forces for Davydov components are exclusively of the vdW interactions. The calculated frequency for Davydov components in NL MoTe$_2$ (N = 3–6) are depicted in Fig. 6(b) by open diamonds and squares. The calculated result based on the vdW model is in good agreement with the experimental data. This further confirms that the observed multiple peaks at ~171 cm$^{-1}$ shown in Fig. 5(a) are Raman-active Davydov components in NL MoTe$_2$ (N > 2).

In principle, the frequency of uncoupled entities in NL MoTe$_2$ should be equal to that of the isolated entity because all the nearest Te atoms in adjacent layers vibrate in phase. Therefore, the frequency of the $A_1'$ (R$_2$) mode in 3L MoTe$_2$ and the $A_1'$ (R$_3$) mode in 5L MoTe$_2$ should be the same as that of the $A_1'$ mode in 1L MoTe$_2$. However, it is not the case in NL MoTe$_2$ because there should exist long-range Coulombic interlayer interactions in NL MoTe$_2$, which are dependent on N, similar to the case in NL MoS$_2$ [31,38,39]. Indeed, the frequency difference between the $A_1'$ mode in 1L MoTe$_2$ and the A 1(R$_3$) mode in 5L MoTe$_2$ is about 2.6 cm$^{-1}$. Even so, the Davydov splitting in each NL MoTe$_2$ can be well understood by the vdW model. This suggests that the frequency difference between Davydov components in NL MoTe$_2$ is mainly determined by the interlayer vdW interactions, which opens the possibility to study the interlayer vdW interactions in other layered materials by Davydov splitting of the high-frequency optical modes.

It is noteworthy that the Davydov components of the $A_1'(A_{1g}^2)$ modes in the present work are more obvious than those reported for MoSe$_2$ and WS$_2$ [37,40]. In principle, similar Davydov splitting of other optical modes in NL MoTe$_2$ can also be observed, e.g., the $E'(E_g^1)$ mode at ~234 cm$^{-1}$ and the $A_2''(A_{1g}^1)$ modes at ~293 cm$^{-1}$. However, it is not the case for the $E'(E_g^1)$ and $A_2''(A_{1g}^1)$ modes. In fact, it is a challenge to simultaneously observe Davydov doublets for NL MX$_2$. For example, Davydov doublets have not been observed in the Raman spectra of few-layer MoS$_2$ and WSe$_2$ so far, and no Davydov splitting of $A_1'(A_{1g}^2)$ modes has been observed in the Raman spectra of NL MoTe$_2$ in the two previous works [14,17]. Figure 4 and a recent work [24] indicate that the observation of the distinct Davydov splitting of the $A_1'(A_{1g}^2)$ modes in NL MoTe$_2$ results from the resonant Raman enhancement of the Raman intensity by the 1.96-eV excitation because it is close to the energy of B′ exciton in $N$L-MoTe$_2$ [14].

**E. Resonant profile of Davydov components in Multilayer MoTe$_2$**

In order to further study the resonant mechanism of Davydov components of the $A_1'(A_{1g}^2)$ modes in $N$L MoTe$_2$, we used ten excitation energies from 1.58 to 2.54 eV to measure the $A_1'$ modes of 3L MoTe$_2$, as shown in Fig. 7(a). Two Raman modes, $A_1'$ (R$_1$) and $A_1'$ (R$_2$), are observed, whose intensities are normalized to the A$_3$ modes in quartz at ~465 cm$^{-1}$ [20,21] to eliminate the difference of CCD efficiencies at different excitation energies [21]. We also measure the reflectance contrast (ΔR/R) spectra of 3L MoTe2 in the visible range, as shown in Fig. 7(b) as the dashed gray line. Based on the previous results [14], the B, A', B', C and D exciton peaks have been assigned in the ΔR/R spectra. The intensity of the $A_1'$(R$_1$) and $A_1'$(R$_2$) peaks as a function of the excitation energy is plotted in Fig. 7(b), respectively. It is obvious that the intensity of the $A_1'$(R$_1$) peak is greatly enhanced at 1.71- and 1.58-eV excitations, which are close to the energy of the A' exciton in 3L MoTe$_2$. The strong $A_1'$(R$_1$) peak resonant with the A' exciton is attributed to the exciton-phonon interactions [20,40,41]. The $A_1'$ (R$_2$) peak shows strong intensity once the excitation energy is near 1.85 eV, which is close to the energy of the A' exciton (~1.73 eV) and B' exciton (~1.96 eV). It indicates that the intensity enhancement of the $A_1'$(R$_2$) peak is mainly resonant with the A' and B' excitons. With respect to the large excitation energy, the frequencies of the $A_1'$ (R$_1$) and $A_1'$ (R$_2$) peaks are almost identical to each other. However, the two modes exhibit different resonant profiles, which may result from the differences in the electron-phonon coupling strength of the two modes. Indeed, similar results have been observed for the different C and LB modes in twisted multilayer graphenes [20,21]. Although our result in Fig. 7(b) shows the resonant enhancement for the $A_1'$(R$_2$) peak when the excitation energy is close to the A' exciton, however, the $A_1'(R_1)$ intensity is strongly enhanced in this excitation energy by the A' exciton, which make it difficult to distinguish the $A_1'(R_2)$ peak from the strong $A_1'(R_1)$ peak for the excitation energies of 1.71 and 1.58 eV, as indicated in Fig. 7(a). Thus, one must choose proper excitation energy to observe Davydov splitting of the $A_1'(A_{1g}^2)$ mode in $N$L-MoTe$_2$ ($N$>2) due to the different resonant profiles between two Davydov components.

## III. CONCLUSIONS

In summary, we have studied the Raman spectra of fewlayer MoTe$_2$ in both ultralow-frequency and high-frequency regions. In the ultralow-frequency region, the frequencies of C and LB modes agree well with the prediction based on the LCM in which only nearest interlayer coupling is considered. The intensity of the lowest-frequency LB mode is much stronger than that of the C mode. This phenomenon is opposite to the reported results for few-layer MoS$_2$ and WSe$_2$. The results indicate that the second-nearest layer-breathing interlayer coupling and the substrate effect can be ignored in the analysis of Raman spectra of exfoliated NL MoTe$_2$ on different substrates. Under resonant excitation conditions, Davydov splitting of the out-of-plane $A_1'(A_{1g}^2)$ modes at ~170 cm$^{-1}$ is observed. The number of the Davydov components and their frequencies are dependent on layer number. Based on the symmetry analysis, all the predicted Raman-active $A_1'(A_{1g}^2)$ modes in NL MoTe$_2$ (N = 3–6) have been assigned. It is noteworthy that the Davydov splitting of the $A_1'(A_{1g}^2)$ modes is more obvious than those reported for MoSe$_2$ and WS$_2$. The Davydov splitting of high-frequency $A_1'(A_{1g}^2)$ modes has been well understood by the van der Waals model, in which only the nearest interlayer coupling is taken into account. The resonant behavior of the $A_1'$ modes in 3L MoTe$_2$ indicates that the difference in the electron-phonon coupling strength between two Davydov components may result in different resonant profiles, and thus proper excitation energy must be chosen to observe the Davydov splitting of the $A_1^{'}(A_{1g}^2)$ modes in NL MoTe$_2$ (N > 2). The detailed exploration for Davydov splitting in few-layer MoTe$_2$ reveals how the van der Waals interactions significantly affect the frequency of the high-frequency intralayer phonon modes and expands our understanding on the lattice vibrations and interlayer coupling of transition metal dichalcogenides. Note added. Recently, we became aware of a preprint reporting resonant Raman scattering in few-layer MoTe$_2$ [42]..

## ACKNOWLEDGMENT

We We acknowledge support from the National Basic Research Program of China (Grants No. 2013CB921901 and No. 2012CB932703) and the National Natural Science Foundation of China (Grants No. 11225421, No. 11434010, No. 11474277, No. 61125402, No. 51172004, and No. 11474007).

Q.J.S. and Q.H.T. contributed equally to this work.

‡ lundai@pku.edu.cn

* phtan@semi.ac.cn

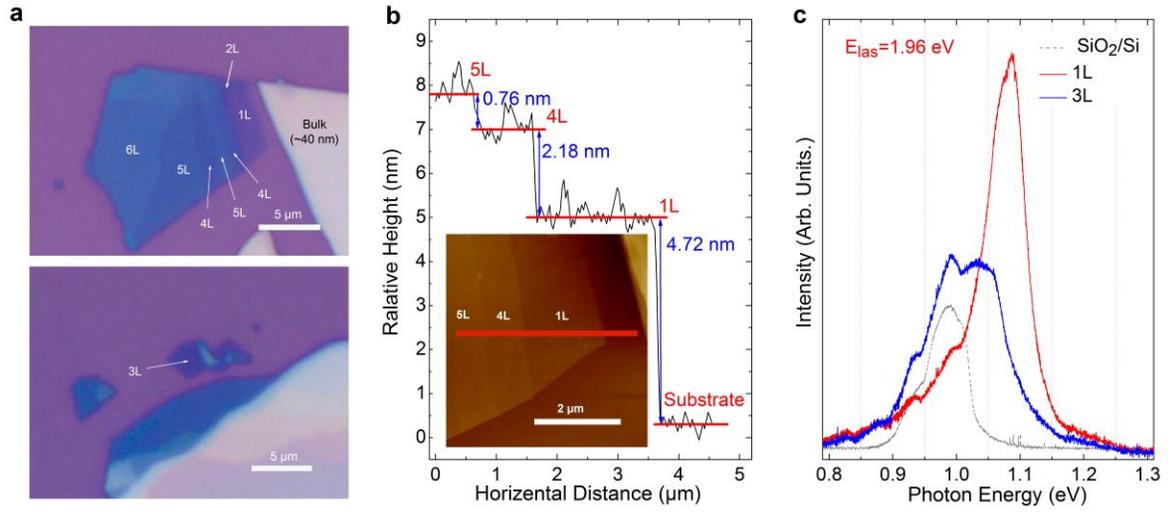

FIG. 1. Optical microscope image of 1L–6L and bulk MoTe2 flakes on 300-nm SiO2/Si substrates. (b) Sample heights along the red line in the inset. Inset: the AFM image of the measured region. (c) PL spectra for 1L and 3L MoTe2 on a 300-nm SiO2/Si substrate under the 1.96-eV laser excitation.

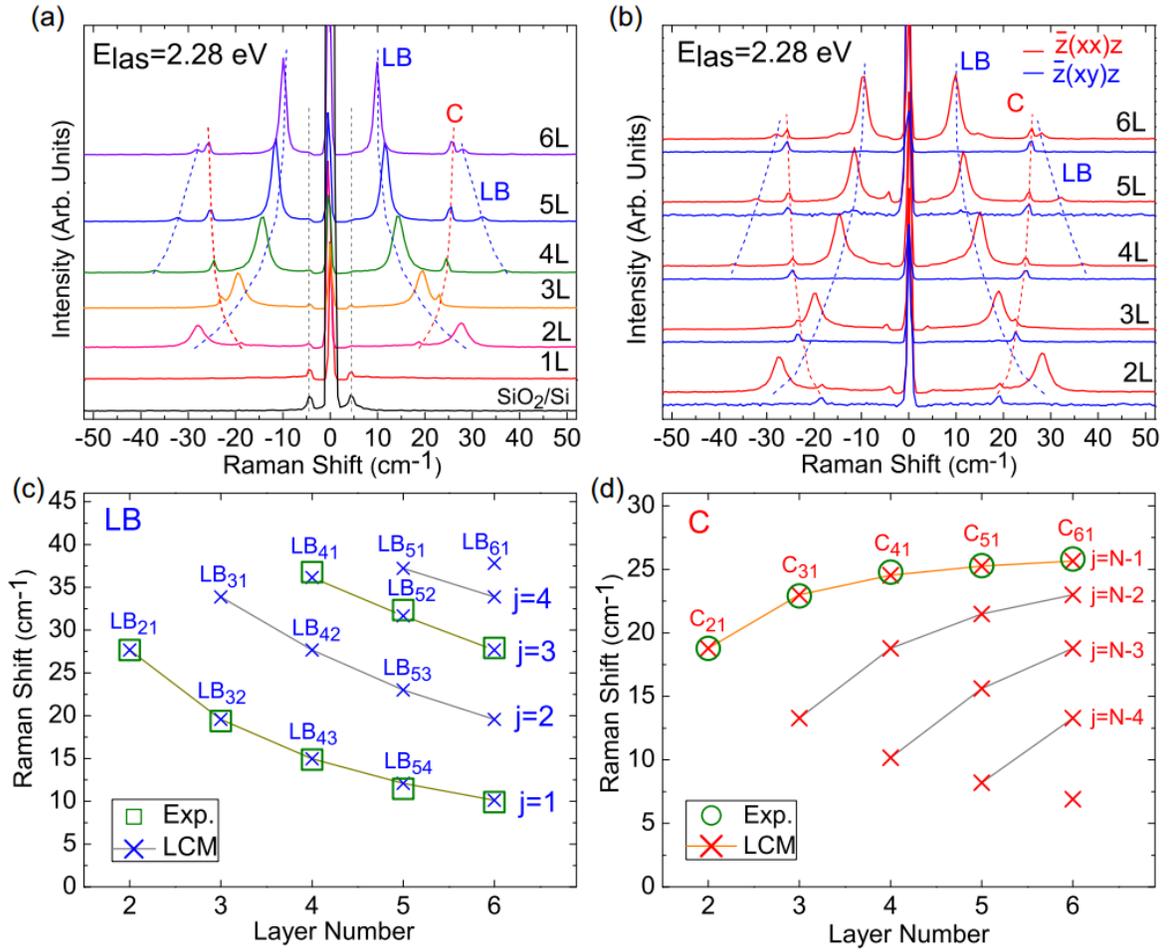

FIG. 2. (a) The Raman spectra of 1L–6L MoTe2 in the ultralow-frequency region. (b) The Raman spectra of 2L–6L MoTe2 under the parallel (red solid line) and perpendicular (blue solid line) polarization configurations. Blue and red dashed lines are used to link the LB and C modes, respectively. (c,d) The experimental frequencies and the calculated ones of the LB and C modes in 2L–6L MoTe2 based on LCM.

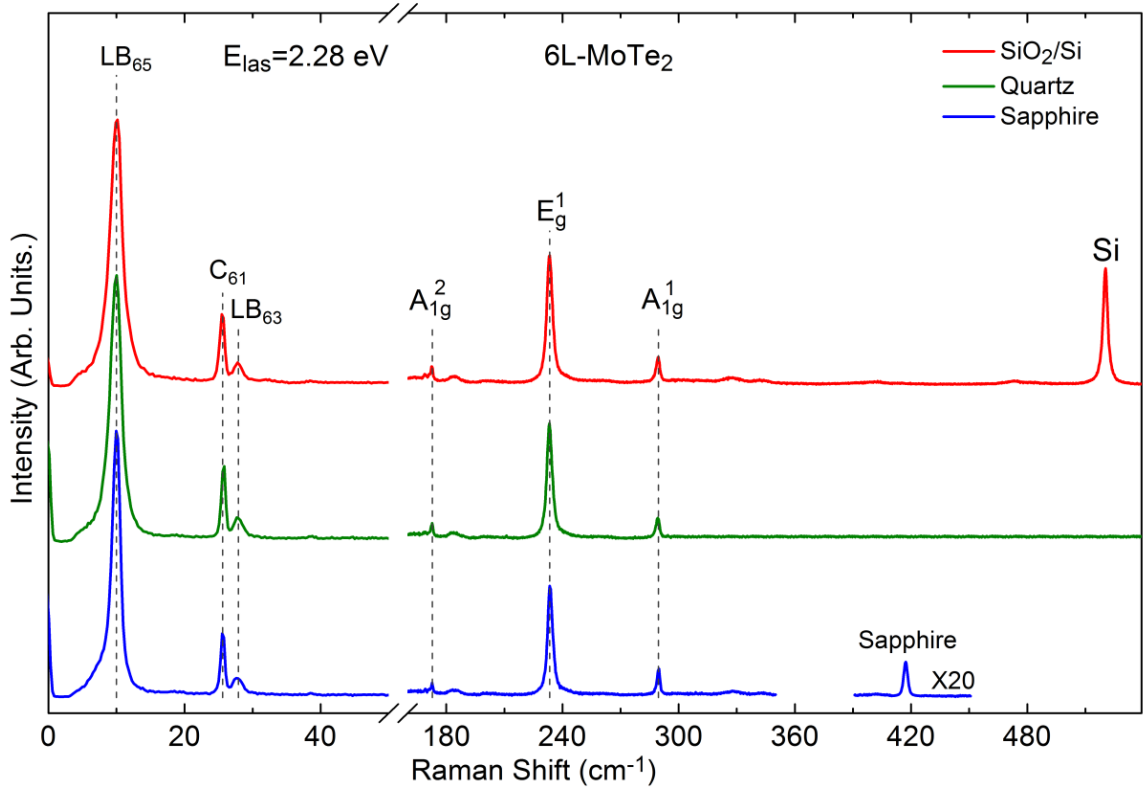

FIG. 3. The Raman spectra of 6L MoTe2 with red, green, and blue lines for samples on SiO2/Si, quartz, and sapphire substrates, respectively. The spectra are normalized to the strongest LB mode located at ~10 cm$^{-1}$.

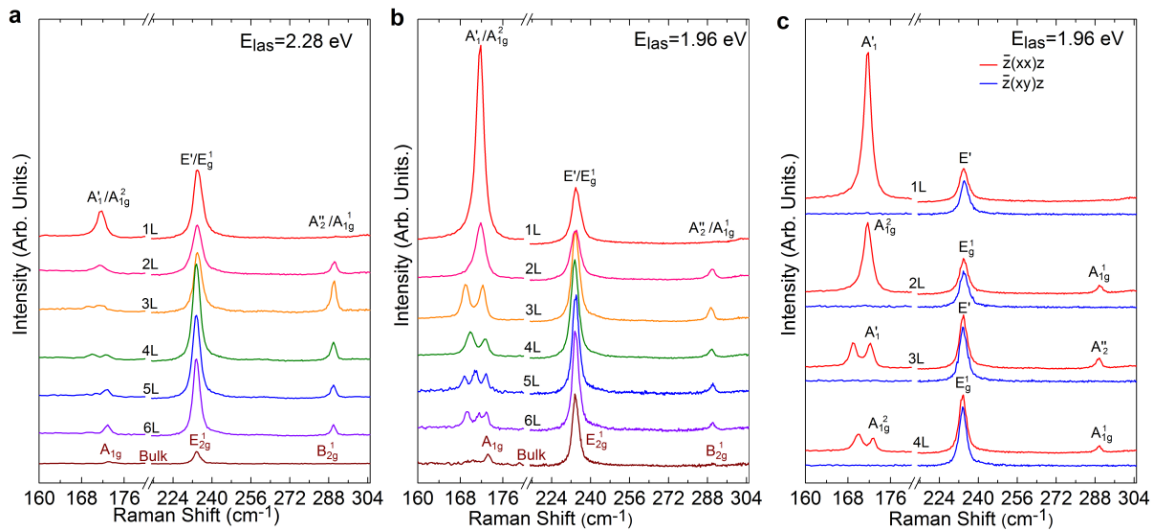

FIG. 4. (a,b) Raman spectra of 1L–6L MoTe2 and bulk one on SiO2/Si in the high-frequency region excited by 2.28 and 1.96 eV, respectively. The corresponding irreducible representation is

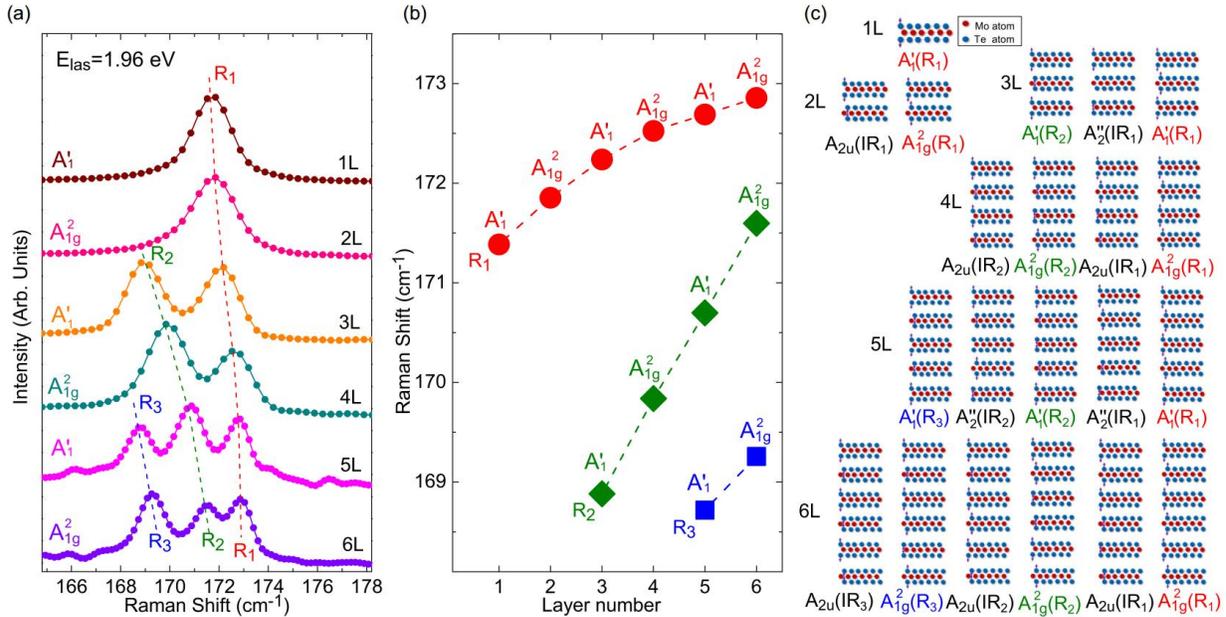

FIG. 5. (a) The $A_1^{'}(A_{1g}^2)$ modes in 1L–6L MoTe2 excited by 1.96 eV. Raman spectra are normalized to the strongest peak and are offset for clarity. (b) The experimental result of the frequency evolution of the $A_1^{'}(A_{1g}^2)$ modes with layer number. (c) Normal atomic displacements for all the high-frequency modes 2L–6L MoTe2 which are derived from the $A_1^{'}$ mode in 1L MoTe2. The relative motion of Te atoms is schematically drawn by the left atoms with purple arrows. The corresponding irreducible is also listed for each mode, where the Rj and IRj (j = 1,2, or 3) in the brackets are used to identify the Raman-active and infrared-active modes for each NL MoTe2. All the modes are arranged in frequency from low to high from left to right.

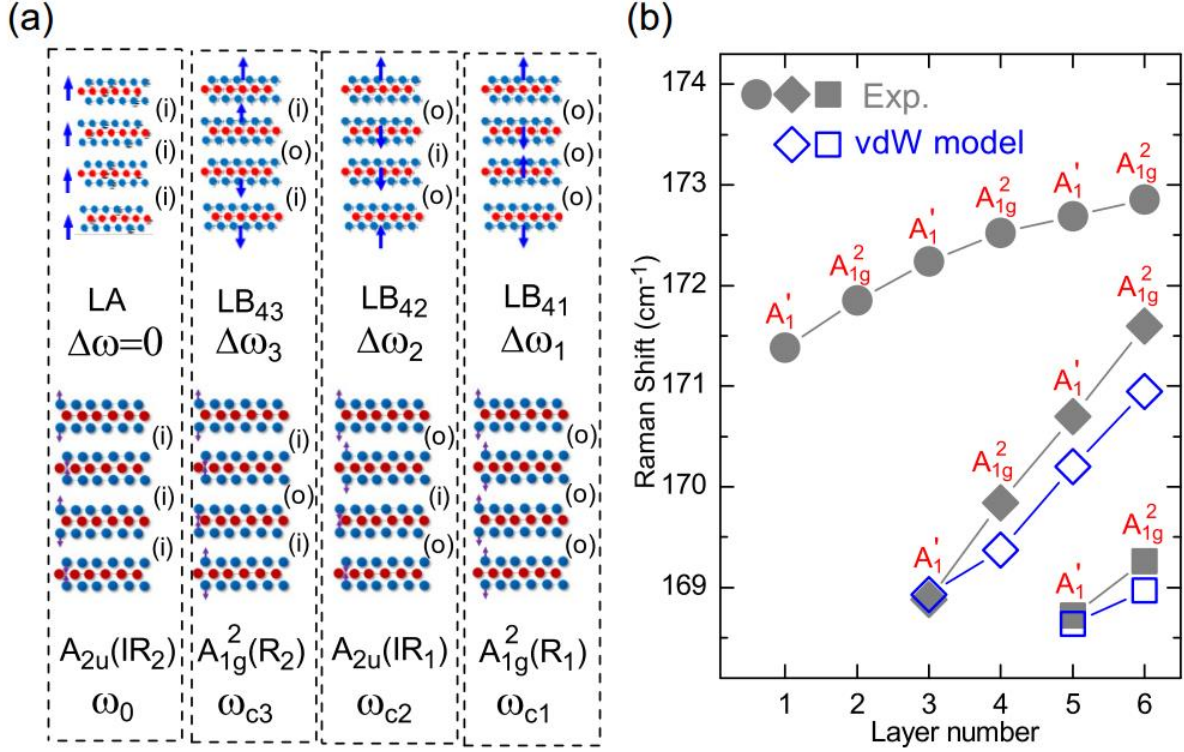

FIG. 6. (a) Schematic diagram of the vdW model for Davydov splitting in 4L MoTe2. Four Davydov components (two $A_{1g}^2$ and two A2u modes, bottom panel) derived from the $A_1'$ mode in 1L MoTe2 and the corresponding four coupling modes (three LB modes and one LA mode, top panel) between four coupled MoTe2 layers are shown. The coupling frequency for the three LB modes is denoted as Δω1, Δω2, and Δω3, respectively. The A2u(IR2) mode is the uncoupled entities with a frequency of ω0, and the other three modes are the coupled entities with the frequencies of ωc1,ωc2, and ωc3, respectively. The (i) and (o) in the each atomic displacement denote in-phase and out-of-phase vibrations of Te atoms in adjacent layers, respectively. The number of out-of-phase vibrations of Te atoms in adjacent layers for each $A_{2g}^1$ or A2u mode is the same as that of the corresponding coupling mode. (b) The calculated frequency (open diamonds and squares) of each Davydov component of the $A_1'(A_{1g}^2)$ modes in 3-6L MoTe2 based on the experimental (Exp.) value (gray solid circles) of the Davydov component with highest frequency and the vdW model of $\omega_{cj}^2 - \Delta\omega_j^2 = \omega_0^2$ (j = 1,2,3), and the corresponding experimental (gray solid diamonds and squares) frequency of each Davydov component of the $A_2''(A_{1g}^1)$ modes in 3-6L MoTe2. The experimental frequency of the $A_1'(A_{1g}^2)$ modes in 1-2L MoTe2 is also included.

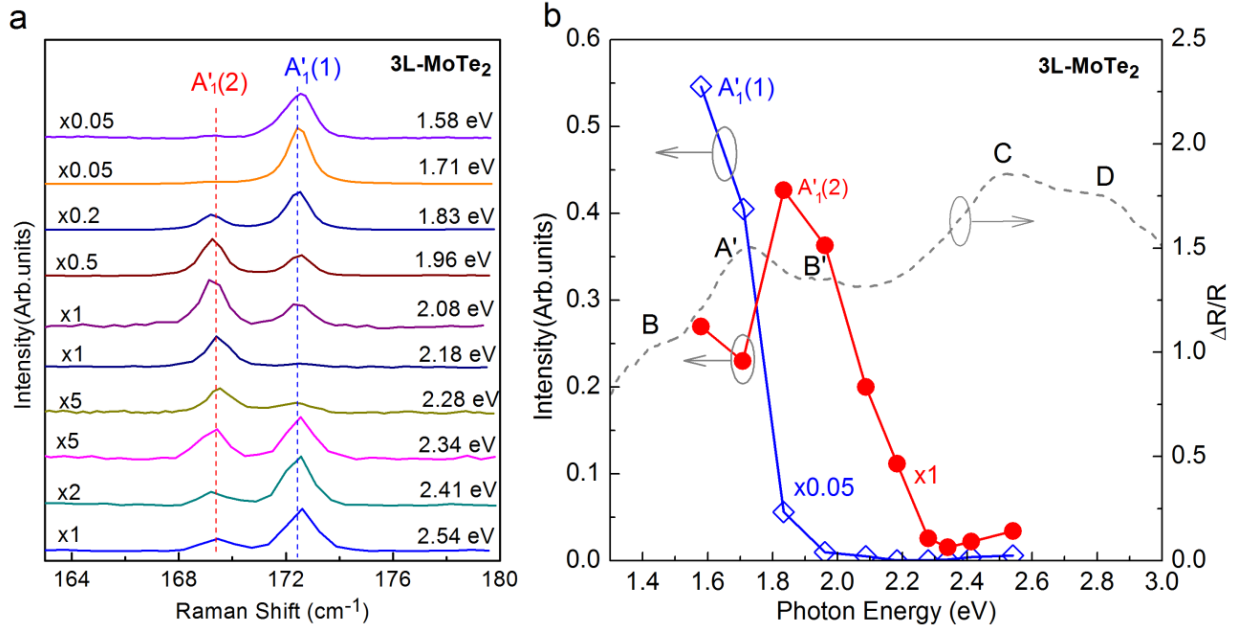

FIG. 7. (a) Raman spectra of Davydov doublets of the $A'_1$ modes in 3L-MoTe$_2$ excited by 10 laser excitation energies, where the Raman intensity is normalized to the A$_3$ mode in quartz at about 465 cm$^{-1}$. (b) The intensity of the $A'_1$(R$_1$) (blue squares) and $A'_1$(R$_2$) (red circles) as a function of the excitation energy. The dashed gray line is the reflectance contrast spectrum ($\Delta$R/R) of 3L-MoTe$_2$ in the visible range.